# SPACECRAFT AUTONOMY LEVELS

Daniel Baker[*] and Sean Phillips[†‡§]

As autonomous systems are being developed in multiple industries it has been recognized that a phased approach is needed both for technical development and user acceptance. Partially in response, the automotive and aircraft industries have published five or six Levels of Autonomy in an attempt to broadly characterize the amount of autonomy technology that is present in any particular vehicle.

We propose six Spacecraft Autonomy Levels to characterize and describe the high-level autonomous capabilities for any given satellite. Our proposed spacecraft autonomy levels parallel those of the automotive and aircraft levels in an attempt to maintain consistency and a common understanding. We see these autonomy levels as useful in education and communication with lawmakers, government officials, and the general public.

## INTRODUCTION

As autonomous systems are being developed in multiple industries, it has been recognized that a phased approach is needed both for technical development and user acceptance. For most manufactures, each new model of automobile introduces new autonomous features and elements (e.g. adaptive cruise control and automatic parking). At some point these individual features combine to form a higher-level capability (e.g. autonomous speed control and autonomous steering control). From early in space flight's history, most spacecraft systems have incorporated automated processes. Some examples include automatic sun pointing if power levels drop below a predefined level; regularly scheduled ground communication contracts; and transition to a safe mode when anomalies occur. In space, the push for more automation is motivated by several factors: the remoteness of the spacecraft to its human controllers, the inability to repair hardware problems, and the communication time delays and time gaps between the spacecraft and the ground to name a few.

In an attempt to broadly characterize the amount of autonomy technology in a particular system, many industries have published five or six levels of autonomy. Figure 1, shows the six SAE Levels of Driving Automation for automotive vehicles and Figure 2, shows five levels of autonomy for aircraft.[13, 2] Similar levels, typically on a 0-5 scale, have been proposed in other industries, including:

- Drone Flight ([5], [8], [14])
- Industrial and Manufacturing ([4])

---

[*]Aerospace GNC Engineer, Defense, BlueHalo, LLC., Albuquerque, NM, USA.
[†]Technical Advisor, Space Control Branch (RVSW), Air Force Research Laboratory, Albuquerque, NM, USA.

- Robotic Surgery ([23])
- Information Technology and Telecommunications Networks ([7], [6]).

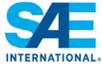

Figure 1. SAE Levels of Driving Automation

We propose a similar six levels (0-5) of spacecraft autonomy. We intend these autonomy levels to characterize and describe high-level capabilities for any given satellite. Defining spacecraft autonomy levels in this way maintains consistency with other domains as autonomy capabilities are increasing added to systems. It is a top-down approach to developing autonomy concepts which can be useful in education and communication with lawmakers, government officials, and the general public. We also see this as a launch point for some common understanding and language for future discussions as we work to get a more integrated and comprehensive bottom-up architecture developed.

The remainder of the paper is organized as follows: The next section summarizes some background material that will be important to the following discussion. Then previous examples of autonomy levels in other industries are reviewed. Then we introduce and discuss our proposed Spacecraft Autonomy Levels. We end with a summary and our conclusions.

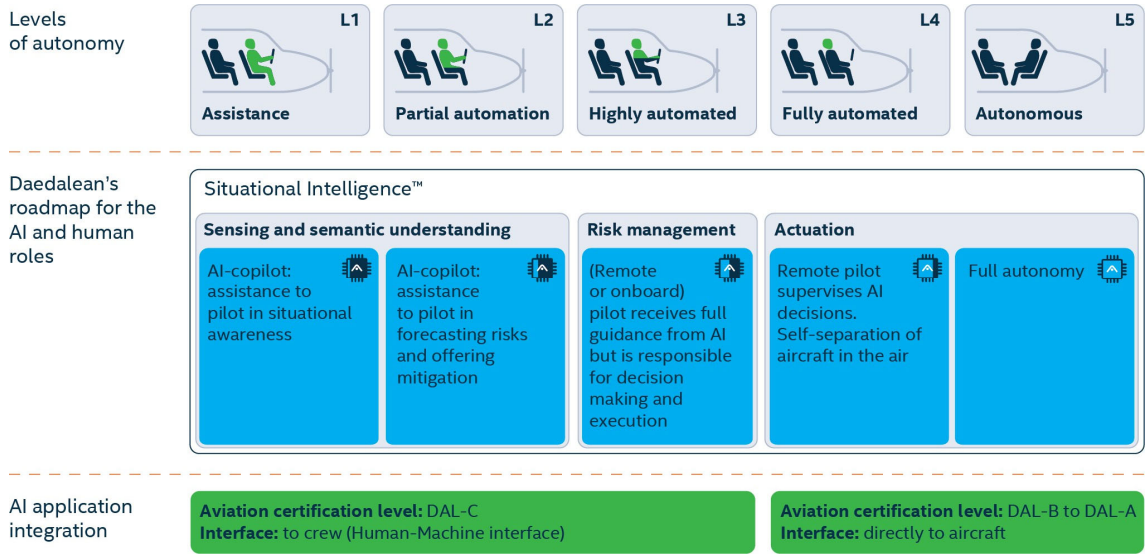

Figure 2. Aircraft Autonomy Levels

**BACKGROUND**

Autonomy means many things to many people. We will not attempt to define autonomy here, but simply recognize that we all need to come to some general understanding of and way of communicating about autonomy. To that end, we are developing a practical framework for autonomy to assist in the design, implementation, use, and test of autonomous systems. We need a usable working description and a way to organize and communicate what, how, and why about the autonomy capability we are discussing.

Autonomy (for humans or machines) invokes the notion of independence. The ability to direct one's own decisions from personal life to interactions with the world. In other words, to control what one does, when one does it, and who one does it with. From the human perspective, it is easy to see this independence play out between children and parents and between workers and managers. For machines, in particular, we primarily think of the system being able to function without (or with little) need for human intervention. Which is back to the basic independence idea. Whether for humans or machines, decision making is a key skill and especially explain-ability of decisions and actions. Rational, understandable, and reasonable decision making is what facilitates trust.

Even though we propose six spacecraft autonomy levels, we see autonomy as a continuous sliding scale from pure automation of repetitive tasks to complete autonomy. This can also be seen as a scale from dependence (like for a new born baby) to independence (of an adult). For machines it looks like a scale from dependence on to the independence from its system design, given knowledge, and a priori knowledge. Figure 3 summarizes some of the aspects of the two extremes of the scale.

In [3], Antsaklis and Rahnama make the argument that "Autonomous systems evolve from control systems ...". We share this viewpoint and come at it from a slightly different perspective. Autonomy is about getting things done. Machines (systems, things, etc.) need to move to do things (work). The motion of satellites, cars, planes, robots, machines, etc. is governed by physics and described by differential equations. Controls is the engineering discipline for regulating motion and we see autonomy as the engineering discipline for regulating controls. That is, autonomy uses controls to execute actions in order to achieve its (autonomy's) goals. Autonomy is the who, what, when, and where is to be done and controls is how it is done. Autonomy determines what needs doing and controls does the doing.

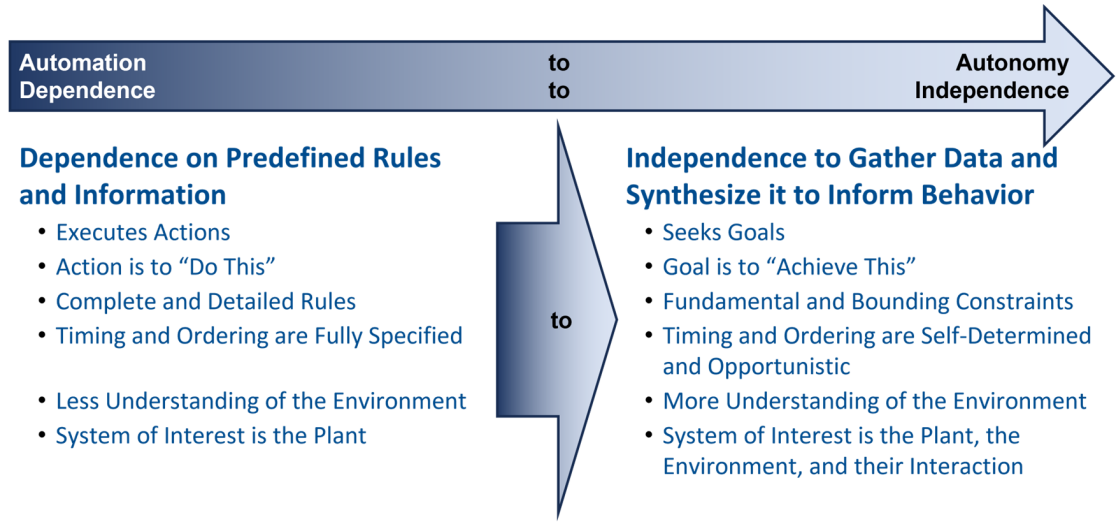

**Figure 3.  Extremes of Autonomy on a Sliding Scale**

The purpose of an autonomous system, in most peoples minds, is to replace the human. Factory robots replace (some of) the humans doing assembly tasks. Household robots do tasks or chores for (replacing) the humans. Basically we want to do through hardware and software what Mother Nature has done through biology. This is a useful analogy. The autonomy aspect of a robotic vacuum cleaner is not to vacuum, the machine already does that. The autonomy aspect of a robotic vacuum cleaner is to replace the human's direction about when and where to vacuum. In this sense the autonomy system acts as a manager, supervisor, director, and monitor. It has decision making authority within its sphere of responsibility; directs activities and tasks; monitors everything, intervenes when needed to meet some goal(s); and provides reporting up to higher levels. The reporting can be about actions taken and/or decisions or actions needed from the higher levels. In our robotic vacuum example, reporting could include when vacuuming was completed or next scheduled to begin. It could also report the need for recharging or emptying of the dirt container (if it is not capable of doing these tasks itself).

We have identified four primary (high-level) functions (or purposes) of all autonomy systems: Survival, Success, Collective, and Contextualization of Situations (the SSCCs). Next, we describe these four purposes.

**Survival** refers to the fact that all autonomy systems want to act in a way that ensures its own continuation. Humans have this innate survival instinct and autonomy systems should as well. Autonomy cannot achieve its goals if it ceases to operate. Survival includes all of our normal ideas of safety, health, and well being as they apply to both humans and machines. In military situations, survival can also imply undetectability and stealth. In a hierarchical autonomy system, a high-level element could allow/mandate the demise (sacrifice) of a lower-level element for survival of larger system. A simple example of the sacrifice of a lower-level element would be of a fuse blowing to protect the larger circuit. In a military context, an example is the sacrifice of a soldier for the good of the larger campaign, even though the solder has high survival motivations.

**Success** refers to the function of the autonomous system to meet or attempt to meet its mission, goals, and objectives. Its aim to achieve this success is in spite of knowing the system constraints, limitations and potential impacts of decisions. Success gets to the core of the idea of replacing the human by accomplishing the tasks that humans would do. This purpose clearly involves the skills of being able to develop plans, priorities, and tasks within given constraints, boundaries, and authority and being able to execute those devised tasks.

**Collective** is the third function of an autonomy system. No autonomous agent (human, system, or machine) operates in complete isolation from other agents. Autonomy must work with other elements and agents and this implies communication among the agents and elements. The working together can serve several purposes: reporting, commanding, and sharing information and/or tasking. With humans in the loop, this also needs to include human-machine interface design and task load concerns.

**Contextualizing Situations** is maybe the most difficult function of the autonomy system. Here, the autonomous system is trying to replace the human capabilities of taking advantages of new opportunities, anomaly and hazard detection, extrapolation of past experiences to new situations, reasoning and understanding, problem solving, and developing plans for action. These situations can stem from changes in the environment, changes in the machine, ineffective previous plans, and unknown and unforeseen circumstances.

**EXAMPLES OF EXISTING AUTONOMY LEVELS**

Against this background, we examine some of the examples of existing autonomy levels.

Starting with the six driving automation levels (Figure 1), the focus is on just the two aspects of driving: speed (acceleration and deceleration) and steering control. There is also no use of the word autonomy. Ignoring these two limitations (for our purposes as related to spacecraft), there are clear distinctions between when the human is responsible for driving (blue blocks) and when the vehicle is responsible (green blocks). Even when the human is responsible for driving (levels 0-2), the graphic defines what kind of assistance (to the human) the automation features provide.

The second example from above (Figure 2), is that of aviation autonomy levels. Level 0 is omitted but can be easily implied as the case where there is no flying assistance or artificial intelligence (AI) capability. It is also interesting to note that the graphic implies that autonomy functions are being provided by some kind of AI. We make no such assumption about an autonomy system. We believe that many autonomy functions can be provided by simpler components than an AI system.

In the aviation autonomy levels graphic, the levels are titled and the titles are fairly consistent when comparing them with other examples. Unlike the driving automation levels, the aviation levels do not limit themselves to speed and steering control. Instead, they focus on a progression of the autonomy system's situational awareness capabilities. Starting with sensing, the levels move through an understanding of the situation to being able to identify potential risks and suggest possible mitigation actions. Then at L5, the autonomy system has the capability to act independently. The graphic also identifies where certification levels map to the autonomy levels which at some point may become important for spacecraft as well.

In other examples of autonomy levels, we continue to see the theme of changing roles between the human and machine. At level zero the human has responsibility for situational awareness, analysis, decisions, and actions. As the level increases the system gradually assumes these roles and the human takes an oversight and possible override role.

In robotic surgery [23], Yang, et al. define six levels spanning from Level 0 where no autonomy exists and the human performs all tasks to Level 5 where a robot can perform the entire surgery and no human is needed. The level distinctions are based on who does and who decides. In autonomy levels for work, [1] and [16] continue the same themes with AI systems gradually performing more tasks and roles. In the telecommunications and information technology [7], [6], the focus is on the transition from human programmed scripts to system self-adaptation for determining and completing tasks. For automating industrial and manufacturing facilities, [4] in their article on Industrial Automation to Industrial Autonomy (IA2IA), they also use six levels that move from human control of operations to autonomous control with little or no human oversight.

Even outside of technical disciplines, the idea of autonomy levels persists. In [12], Homer outlines the five levels an organization autonomy where level 1 organizations need lots of outside control (micromanaged) to level 5 where the organization is self-governed and in alignment with overall goals. For the task of delegation, Tounsi [21] refers to Webber's eight autonomy levels were the levels are differentiated by who does the analysis, who makes the decision, and who takes the action. Note that level 1 is most other's level 0 and level 7 where the delegate only communicates with the manager if the action was not successful is not usually separated out in other levels. In individual human learning, [20], a student moves, in four levels, from outside directed (dependent) learning to self-directed (autonomous) learning.

Getting back to examples closer to space flight, the six drone autonomy levels [5], [8] [14], are very similar to those of aviation autonomy. In [9], Clough proposes 11 levels for uncrewed aerial vehicles (UAVs). The first six levels (0-5) are with respect to a single UAV and levels 6-10 are for clusters of UAVs. The single vehicle levels and descriptions are similar to the drone levels. Additionally, Clough adds the elements of perception/situational awareness, analysis/decision making, and communication/cooperation in their initial levels and then in their final levels changes these elements to the observe, orient, decide, and act (OODA) loop aspects.

The National Institute of Standards and Technology (NIST) has also produced a publication [18] on autonomy levels for unmanned systems. While applicable to spacecraft, it is predominately geared towards ground, sea, and air systems. They define 11 levels (0-10) but in looking at their graphic (Figure 4) of the levels, they make no distinction between levels 1-2, 4-5, and 7-9. The result is a five level scheme.

The North Atlantic Treaty Organization's (NATO) Industrial Advisory Group (NAIG) in [15] define four levels. Most notably, they distinguish between an autonomous non-learning system (level 3) and an autonomous learning system (level 4).

In the space domain, the European Space Agency (ESA) has identified four mission execution autonomy levels [17] and Williamson, et. al. [22] have identified 9 levels for small satellite clusters. A recent National Aeronautics and Space Administration (NASA) report, [19], identified seven levels of navigation autonomy. Their levels are very similar to what we will propose below with the exception that their level 0 (no autonomy) does not exist in in our six levels.

In summary, consistent themes emerge from these examples. All show a progression from the system's dependence on human input and/or direction to independence. The autonomous system gradually gains authority. By authority, we mean the ability to control its analysis, organizing, deciding, and acting. The who, what, when, where, why, and how, it does its work. Many of the example levels explicitly discuss transitions in who has decision making capabilities and responsibilities. Some of the examples are also explicit in identifying who actually takes the action(s) at each level. Some examples also make explicit the ideas of learning or self-adaptation. All of the examples implicitly assume that the autonomous system will be replacing the human in all aspects except for possibly oversight.

The other element that is not explicit in any of the example levels is the aspect of complexity. For autonomy, the question and impact of complexity affects three aspects: the system, the environment, and the tasks, goals, or missions. At lower autonomy levels, we might expect to see lower complexity in at least one of these aspects. At higher autonomy levels, we might expect to see the complexity of each of these aspects rise. We will follow the lead of these other examples and ignore the issue of complexity for now. Our future work will expand to include complexity.

The final work to consider is the Space Trusted Autonomy Readiness Levels by Hobbs, et. al. in [10] (see also [11]). They are focused on Technology Readiness Levels which are intended to provide an assessment of a technology's maturity on a scale of 1-9 with 1 being least mature and 9 being the most mature. The work produced both nine autonomy readiness levels (across four categories) and nine trust readiness levels. They do mention autonomy levels, as we have been reviewing here, and conclude that:

> "Conversations about the technical maturity of autonomous space systems inevitably lead to a discussion about levels of autonomy themselves.    and these differ from readiness levels. Autonomy levels usually define specific roles for humans versus automation/autonomy, while readiness levels define the technical maturity of those systems."

**SPACECRAFT AUTONOMY LEVELS**

Hobbs et al. also states that autonomy levels "do not themselves address the focus of their paper, namely issues of trust and readiness" and thus we propose a now familiar six levels (0-5) of spacecraft autonomy (Table 1 and Figure 5). These spacecraft autonomy levels are intended to characterize and describe high-level capabilities for any given satellite. The spacecraft levels parallel those of the other example levels (reviewed above) in an attempt to maintain consistency and a common understanding with our intended audiences.

In following the examples discussed above, we distinguish between roles of the autonomy system with those of the ground-based human flight controllers (also called operators and mission planners). The human operator's role is to plan, schedule, coordinate, and monitor the activities of some desired mission or maneuver for the spacecraft. If the autonomy system is to gradually replace the human then it is to subsume these roles. Also, frequently, a first level breakdown of a spacecraft's systems separates the bus from the payload(s). When we speak of the onboard spacecraft autonomy system, we are referring to the combined capabilities of the bus and payload(s), i.e. the complete spacecraft.

**Level 0 – Basic Spacecraft Controllability**

We begin with Level 0 as a basic functional, controllable and command-able, spacecraft. At Level 0, the only autonomous functions are the spacecraft's systems being able to trigger a transition to safe mode upon the detection of some anomaly or out-of-bounds variable. All spacecraft situa-

### Table 1. Proposed Spacecraft Autonomy Levels

**Level 0 – Basic Spacecraft Controllability**
- Spacecraft has All Basic Systems Operational
- Spacecraft Autonomy Limited to Automatic Transitions to Safe Modes when Anomalies Occur
- Ground Provides Situational Awareness Analysis and Determines all Spacecraft Actions
- Ground Schedules and Initiates All Communications

**Level 1 – Ground Systems Assistance**
- Spacecraft Onboard Autonomy Keeps Ground Systems Updated on its Local (Internal and External) Situational Awareness
- No (or very little) Ability to Act Independently
- Other Telemetry used by Ground to Confirm the Spacecraft's Situational Analysis

**Level 2 – Advanced Assistance**
- Onboard Autonomy Forecasts Risks and Devises Mitigation Actions in Assistance to Ground Systems
- Still No Authority to Act on Its Own, Initiation by Ground Systems
- Adds Capabilities in Reasoning and Planning (over Level 1)

**Level 3 – Partial Automation**
- Autonomy now Capable of Acting on its Faults and Risks Mitigation Plans Without Ground Approval
- Onboard Autonomy Forecasts Mission Needs and Devises Mission Scenarios in Assistance to Ground Systems
- Autonomy Develops Mission-level Course(s) of Action and Presents them to the Ground for Approval to Act

**Level 4 – Full Automation**
- Autonomy Develops and Executes Mission-level Course(s) of Action and Reports Actions and Results to the Ground
- Ground Systems Monitors Autonomy Decisions and Actions on a Regular Schedule
- Autonomy Collaborates with Ground Systems in Developing Mission Parameters and Mission Planning

**Level 5 – Autonomous**
- Autonomy Develops and Executes Mission Parameters, Scenarios, Plans, and Full Course(s) of Action (reporting available on request)
- Ground Systems Occasionally Monitors Autonomy Decisions and Actions as Oversight

tional analysis (internal and external) and problem solving is done on the ground. The spacecraft is completely commanded by the ground.

**Level 1 – Ground Systems Assistance**

In Level 1, the spacecraft autonomy software has very little ability to act. The autonomy system's function is focused on developing and monitoring its local (internal and external) situational awareness. The spacecraft keeps the ground station updated on its situational awareness. The operators can use other telemetry to confirm that the spacecraft's situational analysis is consistent with their own. The spacecraft is still completely commanded by the ground.

**Level 2 – Advanced Assistance**

At Level 2 the onboard autonomy system is capable of identifying faults and forecasting risks and devising mitigation plans and actions to address the faults and risks. The autonomy system still has little authority for acting. Its additional capabilities, beyond those of Level 1, lie in reasoning and planning. The mitigation plans can be submitted to the ground system for evaluation, modification, and approval. Any action, beyond Level 1, must be initiated by the operator.

**Level 3 – Partial Automation**

A Level 3 autonomy system is now capable of acting on its faults and risks mitigation plans without operator approval. Onboard autonomy can now also forecast mission needs and devises mission scenarios in assistance to ground systems. The autonomy system develops mission-level course(s) of action and presents them to the ground for approval to act. The ground operators monitor the autonomy systems actions and uses the system's mission needs and scenarios as mission planning inputs.

**Level 4 – Full Automation**

At Level 4, spacecraft autonomy develops and executes mission-level course(s) of action and reports its actions and results to the ground. The operators monitor these autonomy decisions and actions on a regular basis. Spacecraft autonomy also now can have the capability to collaborate with ground systems in developing mission parameters and mission planning for new missions and new requirements.

**Level 5 – Autonomous**

Finally at Level 5, the spacecraft develops and executes its own mission parameters, scenarios, plans, and course(s) of action and reports on its activities as requested. Ground systems may occasionally monitor onboard decisions and actions in an oversight role.

An interesting thing to note here is that communication between the spacecraft and the ground changes as the levels change. At the lower levels, the language is very prescriptive detailing the commands and telemetry at an almost raw data level. As the levels move higher the vocabulary is more focused on awareness, tasks, and goals ending with a focus on strategies and tactics.

**SUMMARY**

We have shown how six autonomy levels can be developed for spacecraft as they have been in many other industries. We have attempted to maintain some consistency in level labels and discriminators with these previous examples. We expect these levels to be useful in public relations and marketing materials to be able to provide a basic sense of the capabilities of a particular satellite system. We recognize that the level descriptions (in any of the examples) contain undefined and flexibly interpreted terms but it is a place to start. It also provides little help to designers or builders of autonomy functionality for spacecraft (or any other system, for that matter). With the lack of some ability to quantify the autonomy of a spacecraft from a bottoms-up approach, we propose this top-down representation. We hope it will provide a point of departure for future discussions about the specific automated elements that might be required to claim a given level of autonomy.

It is this future discussion we hope this work will stimulate. What autonomy elements are required for all/most spacecraft and ground stations? Are there autonomy functions that could be optional? Should governments become involved in regulating minimum requirements and certifications? What operations and/or missions are enhanced (increases in effectiveness and/or performance) with autonomy? What operations and/or missions are enabled (otherwise not possible) with autonomy?

Our future work includes a bottoms-up approach to developing autonomous capabilities at all levels within all of a spacecraft's systems and subsystems.

| reference levels: | reference metrics summaries | | | ALFUS Levels |
| --- | --- | --- | --- | --- |
| | MC | EC | HI | User-defined levels using metrics summaries to the left |
| 10 | highest adaptation, decision space, team of teams collaborative missions; fully real-time planning; omniscient, highest level fidelity SA; human level performance | lowest solution/possibility ratio: lowest margin for error, understandability; highest level of dynamics, variation, risks, uncertainty, mechanical constraints* | performing on its own and approaching zero human interaction, negotiating with appropriate individuals | |
| 9 | high adaptation, decision space, team collaborative missions/tasks; high real-time planning; strategic level, high fidelity SA | low solution/possibility ratio, understandability highly dynamic, complex, adversarial high risks, uncertainty, constraints* | UMS informs humans; human provides strategic goals, interacting time between 6 % and 35 %; | |
| 8 | | | | |
| 7 | | | | |
| 6 | limited adaptation, decision space, vehicle tasking; limited real-time planning; tactical level, mid fidelity SA | mid solution/possibility ratio, understandability dynamic, simple mid risks, uncertainty, constraints* | human approves decisions, provides tactical goals, interacting time between 36 % and 65 % | |
| 5 | | | | |
| 4 | | | | |
| 3 | subsystem tasks/skills; internal, low fidelity SA | high solution/possibility ratio, understandability static, simple low risks, uncertainty, constraints* | human decides, provides waypoints, interacting time between 66 % & 95 % | |
| 2 | | | | |
| 1 | | | | |
| 0 | simplest, binary tasks | static, simple | remote control | |

**Figure 4. NIST's Autonomy Levels for Unmanned Systems**

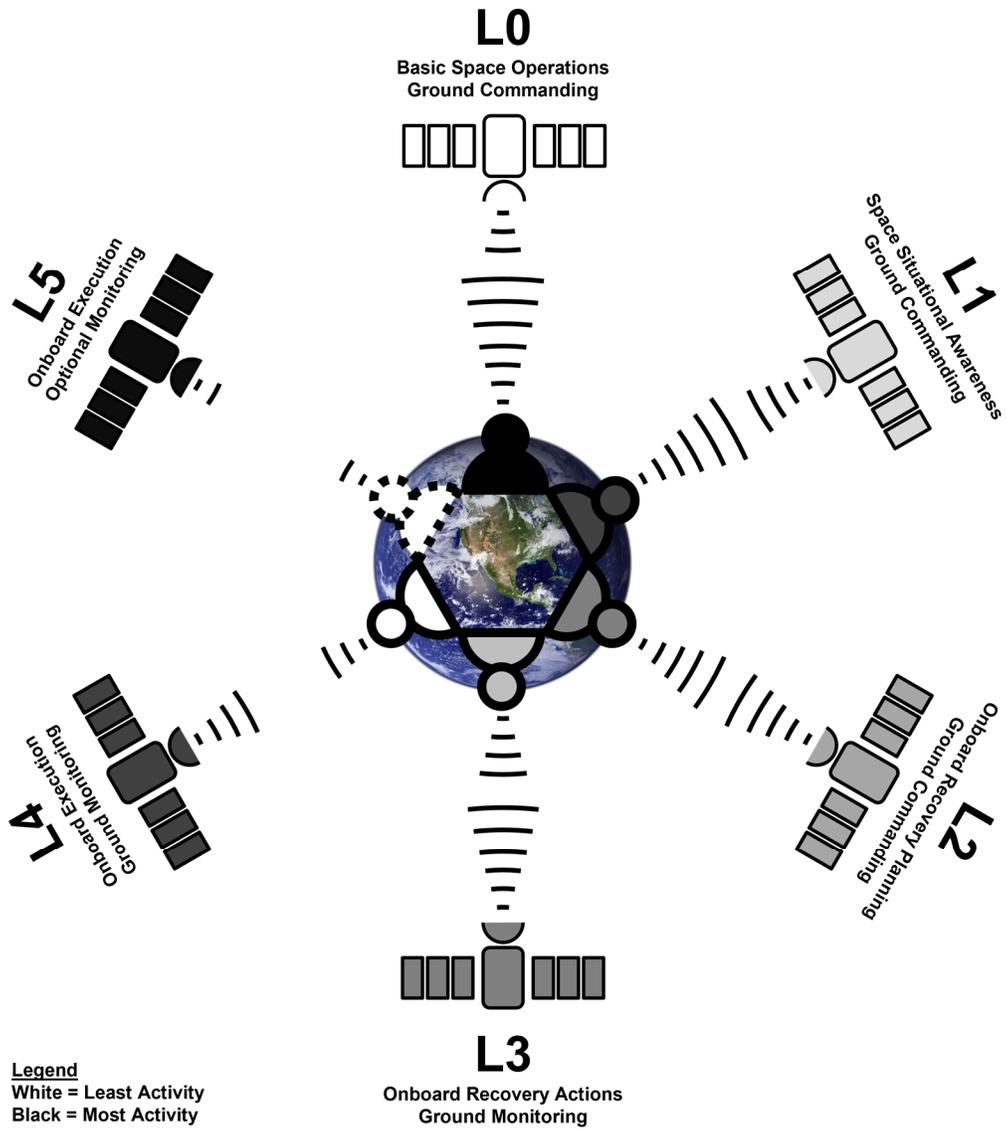

**Figure 5. Spacecraft Autonomy Levels**